\documentclass[journal,10pt,twocolumn]{IEEEtran}
\IEEEoverridecommandlockouts

\usepackage{cite}
\usepackage{amsmath,amssymb,amsfonts,float,graphicx,steinmetz,bm,subfigure,multicol,multirow,physics,mathtools,multicol,multirow,url,color,xcolor,etoolbox,ifthen,cite,hyperref}
\usepackage{textcomp}
\usepackage{tikz}

\newcommand{\bi}{\begin{itemize}}
\newcommand{\ei}{\end{itemize}}

\newcommand{\be}{\begin{enumerate}}
\newcommand{\ee}{\end{enumerate}}
\newcommand{\bd}{\begin{description}}
\newcommand{\ed}{\end{description}}
\newcommand{\bc}{\begin{center}}
\newcommand{\ec}{\end{center}}
\newcommand{\bt}{\begin{tabbing}}
\newcommand{\et}{\end{tabbing}}
\newcommand{\bfig}{\begin{figure}}
\newcommand{\efig}{\end{figure}}
\newcommand{\beq}{\begin{equation}}
\newcommand{\beqarr}{\begin{eqnarray}}
\newcommand{\beqarrn}{\begin{eqnarray*}}
\newcommand{\eeq}{\end{equation}}
\newcommand{\eeqarr}{\end{eqnarray}}
\newcommand{\eeqarrn}{\end{eqnarray*}}
\newcommand{\bflr}{\begin{flushright}\vspace{-0.2in}}
\newcommand{\eflr}{\end{flushright}}
\newcommand{\bsub}{\begin{subequations}}
\newcommand{\esub}{\end{subequations}}
\newcommand{\barr}{\begin{array}}
\newcommand{\earr}{\end{array}}
\newcommand{\nn}{\nonumber}

\def\undb#1{\mbox{\bf{#1}}}

\def\dsp{\displaystyle}

\def\dn{\stackrel{\scriptscriptstyle \triangle}{=}}

\definecolor{lime}{HTML}{A6CE39}
\DeclareRobustCommand{\orcidicon}{%
    \begin{tikzpicture}
    \draw[lime, fill=lime] (0,0)
    circle [radius=0.16]
    node[white] {{\fontfamily{qag}\selectfont \tiny ID}};
    \draw[white, fill=white] (-0.0625,0.095)
    circle [radius=0.007];
    \end{tikzpicture}
    \hspace{-2mm}
}

\foreach \x in {A, ..., Z}{%
    \expandafter\xdef\csname orcid\x\endcsname{\noexpand\href{https://orcid.org/\csname orcidauthor\x\endcsname}{\noexpand\orcidicon}}
}

\begin{document}
\markboth{  This work has been submitted to the IEEE for possible publication. Copyright may be transferred without notice.}{}
\title{ \huge
MIMO Terahertz Quantum Key Distribution
\thanks{This work was supported by the Hong Kong Research Grants Council (grant
number C6012-20G). 
}
\vspace{-0.1in}
}

\author{
Neel Kanth Kundu\orcidA{}, {\em Graduate Student Member, IEEE}, Soumya P. Dash\orcidC{}, {\em Member, IEEE}, \\
Matthew R. McKay\orcidB{}, {\em Fellow, IEEE}, and Ranjan K. Mallik\orcidD{}, {\em Fellow, IEEE}
\thanks{
N. K. Kundu and M. R. McKay are with the Department of
Electronic and Computer Engineering, The Hong Kong University of
Science and Technology, Clear Water Bay, Kowloon, Hong Kong
(e-mail: nkkundu@connect.ust.hk,  m.mckay@ust.hk).

S. P. Dash is with the School of Electrical Sciences, Indian Institute of Technology Bhubaneswar, Odisha, India (e-mail:  soumyapdashiitbbs@gmail.com).

R. K. Mallik is with the Department of Electrical Engineering, Indian Institute of Technology Delhi, New Delhi, India (e-mail: rkmallik@ee.iitd.ac.in). }
\vspace{-0.4in}
}



\maketitle

\begin{abstract}
We propose a multiple-input multiple-output (MIMO) quantum key
distribution (QKD) scheme for terahertz (THz) frequency
applications operating at room temperature. {\color{black} Motivated
by classical MIMO communications, a transmit-receive beamforming scheme is proposed that converts the rank-$r$
MIMO channel between Alice and Bob into $r$ parallel lossy quantum
channels. 
Compared with existing single-antenna QKD schemes, we demonstrate that the MIMO QKD scheme leads to performance improvements by increasing the secret key rate and extending the transmission distance}. Our simulation results show that multiple
antennas are necessary to overcome the high free-space path loss
at THz frequencies. {\color{black} We demonstrate a non-monotonic relation between performance and frequency, and reveal that positive key rates are achievable in the $10-30$ THz frequency range. The proposed scheme can be used for both indoor and outdoor QKD applications for beyond fifth generation ultra-secure
wireless communications systems.}
\end{abstract}



\IEEEpeerreviewmaketitle
\vspace{-0.3cm}
\section{Introduction}
With the exponential increase in demand for high data rates, various physical layer solutions, including multiple-input multiple-output (MIMO) systems and the utilization
of the terahertz (THz) frequency band have been proposed for beyond fifth generation (B5G) applications \cite{faisal2020ultramassive}.

In addition to the requirement of a high data rate, future communication systems are expected to deliver secure and encrypted data transmission to improve reliability and data privacy. In this regard, quantum key distribution (QKD) has
been proposed to meet the challenges of developing highly
secure communication links \cite{manzalini2020quantum}. Current encryption algorithms like Rivest-Shamir-Adleman (RSA), whose security is based on a computationally hard problem (for example, a large integer factoring problem), can be broken by a near-term quantum computer by using Shor's factoring algorithm \cite{manzalini2020quantum,weedbrook2012gaussian}. On the other hand, the security of QKD is guaranteed by the laws of quantum physics that can provide unconditional security \cite{weedbrook2012gaussian}. In general,
the types of QKD studied in the literature are broadly classified into the discrete variable
QKD (DVQKD) and the continuous variable QKD (CVQKD). {\color{black}While DVQKD requires single photon sources and detectors, CVQKD can be implemented by using standard telecommunication equipment \cite{weedbrook2012gaussian}, making it a natural choice for integrating QKD in future wireless networks \cite{ottaviani2020terahertz,he2020indoor,liu2018practical}.}


Recent works have studied the feasibility of CVQKD for next generation wireless communication systems operating at microwave and THz frequencies \cite{he2020indoor,ottaviani2020terahertz,liu2018practical,weedbrook2010quantum}. The asymptotic secret key rates of a single-input single-output (SISO) CVQKD system operating at room temperature in the THz band was characterized in \cite{ottaviani2020terahertz}, which was further extended for an inter-satellite THz link operating at cryogenic temperature in \cite{wang2019inter}. A ray-tracing based SISO channel model for indoor THz CVQKD systems was developed in \cite{he2020indoor}. The authors of \cite{gabay2006quantum} have proposed a MIMO architecture to improve the secret key rates of DVQKD in a free-space optical communication system. Other related works have also proposed the applicability of MIMO architectures for improving the target detection rate and the bit error rate in quantum radar and quantum back-scatter communications respectively \cite{lanzagorta2018virtual,jantti2017multiantenna}.



Here, we propose a MIMO CVQKD scheme to improve the secret key
rate and the maximum transmission distance of a CVQKD system
operating at THz frequencies. We consider a wireless communication
link operating at room temperature between Alice and Bob, which
utilizes CVQKD technology to establish a secret key between them
for securing their data transmission. {\color{black} The choice of THz frequencies is motivated by the fact that the preparation thermal noise varies inversely with frequency. At room temperature, this noise is very high at lower microwave and millimeter
wave frequency ranges, which in turn can make it infeasible to obtain positive secret key
rates \cite{ottaviani2020terahertz,wang2019inter}. } In prior works on THz
QKD, two types of systems have been proposed: one that requires a
THz-to-optical converter for homodyne detection
\cite{ottaviani2020terahertz}; the other that directly uses
microwave/THz homodyne detectors
\cite{liu2018practical,weedbrook2010quantum}. In this work, we
consider the latter system since the THz-to-optical converters
have low efficiency that limits the secret key rates
\cite{ottaviani2020terahertz}. The main contributions of the paper
can be summarized as follows:
\begin{itemize}
\item  {\color{black} We propose a MIMO CVQKD system model and a  a singular-value decomposition (SVD) based transmit-receive beamforming scheme.} This model generalizes the SISO beam-splitter channel model for the THz MIMO channel by utilizing the MIMO quantum back-scatter communication framework of \cite{jantti2017multiantenna}.
\item {\color{black} We present new analytical expressions for the
secret key rate of the proposed MIMO QKD scheme. This demonstrates the non-trivial dependence of performance on the operating frequency, which we show is non-monotonic.  Our analysis also provides a simple necessary condition on the operating frequency in order to obtain non-zero secrecy rate.}
\item We use simulations to compare the performance of MIMO QKD scheme with the baseline SISO scheme. {\color{black} Our numerical results reveal that multiple antennas are necessary to overcome the high free-space path loss in the THz frequency range - an important feature not explicitly considered in earlier works on THz QKD \cite{ottaviani2020terahertz,liu2018practical}. We also show that positive secret key rates are achievable in the 10-30 THz frequency range.}

\end{itemize}
{\em Notation}: Matrices are represented by boldface ($\undb{A}$) letters. Conjugate transpose and transpose of a matrix $\undb{A}$ are denoted by $\undb{A}^{\dagger}$ and $\undb{A}^{T}$  respectively. $\bm{1}_{M\times N} \,, \bm{0}_{M\times N} \in {\mathbb C}^{M\times N}$ represent a matrix of all ones and all zeros respectively, $\bm{I}_M$ denotes a $M\times M$ identity matrix, and ${\rm diag}(\bm{a})$ with $\bm{a} \in {\mathbb C}^{M}$ returns a $M\times M$ diagonal matrix with $\bm{a}$ on its diagonal. $\mathcal{N}\left(\bm{\mu},\bm{\Sigma} \right)$ denotes the real multivariate Gaussian distribution with mean vector $\bm{\mu}$ and covariance matrix $\bm{\Sigma}$.
\vspace{-0.2cm}
\section{System Model}
In the quantum communication regime, the electromagnetic field generated by a transmit antenna oscillating at an angular frequency $\omega$ is quantized, and the Hamiltonian of the system has the same form as that of a quantum harmonic oscillator with $H=\hbar \omega (\hat{a}^{\dagger} \hat{a} + 1/2)$, where $\hbar$ is the reduced Planck's constant, and $\hat{a}, \hat{a}^{\dagger}$ denote the annihilation and creation operator of the harmonic oscillator respectively \cite{helstrom1976quantum,weedbrook2012gaussian}. The quadrature field operators given by $\hat{q}=\hat{a}+\hat{a}^{\dagger}$ and $\hat{p}=i(\hat{a}^{\dagger}-\hat{a})$ are dimensionless canonical observables of the system that are similar to the position and momentum operator of the quantum harmonic oscillator \cite{weedbrook2012gaussian}.  Coherent states are the eigenstates of the annihilation operator $\hat{a}$, given by $\hat{a}\ket{\alpha} = \alpha \ket{\alpha}$, where $\alpha \in \mathbb{C}$ denotes the amplitude of the coherent state \cite{weedbrook2012gaussian}.

We consider a THz wireless CVQKD scheme where Alice and Bob want to establish a secret key between them. We assume that Alice has a transmitter with $N_t$ antennas and Bob has a receiver with $N_r$ antennas. We consider a Gaussian modulated CVQKD scheme where Alice first draws two independent random vectors $\bm{Q}_A, \bm{P}_A\sim \mathcal{N}\left(\bm{0}_{N_t \times 1},V_s\bm{I}_{N_t} \right)$, where $V_s$ is the variance of the initial signal encoding. Alice uses $\bm{Q}_A, \bm{P}_A$ to prepare $N_t$ coherent states $\ket{\alpha_i}$ with $\alpha_i = Q_{A,i}+jP_{A,i} \,,\, \forall \, i=1,2,\ldots,N_t$, and then transmits them from the $N_t$ antennas. Let $\undb{H} \in \mathbb{C}^{N_r \times N_t}$ be the channel between Alice and Bob which is modelled as \cite{busari2019terahertz,deng2014mm}
\begin{equation}
  \undb{H} = \sum_{l=1}^{L} \sqrt{\gamma_l}  e^{j2\pi f_c \tau_l} \bm{\psi}_{N_r}\left(\phi_l^r\right) \bm{\psi}_{N_t}^{\dagger}\left(\phi_l^t\right) \;,
  \label{sys1}
\end{equation}
where $f_c$ is the carrier frequency, $L$ is the total number of multipath components, and $\tau_l$, $\gamma_l$ are the propagation delay and path loss of the $l$-th multipath, respectively. Further, $\phi_l^r$ is the angle of arrival (AoA) of the $l$-th multipath at Bob's uniform linear array (ULA) and $\phi_l^t$ is the angle of departure (AoD) of the $l$-th multipath from Alice's ULA. The array response vector $\bm{\psi}_K\left(\theta\right) $ of a ULA with $K$ antenna elements is given by
\begin{equation}
 \bm{\psi}_K\left(\theta\right) = \frac{1}{\sqrt{K}} \left[1, e^{j\frac{2\pi}{\lambda}d_a\sin\theta},\ldots, e^{j\frac{2\pi}{\lambda}d_a(K-1)\sin\theta} \right]^T 
 \label{sys2}
\end{equation}
where $d_a $ is the inter-antenna spacing, and $\lambda$ is the wavelength of the carrier signal. In (\ref{sys1}), $l=1$ corresponds to the line-of-sight (LoS) path and $l>1$ denote the non-LoS (NLoS) multipath components. The path losses are modelled as \cite{he2020indoor}
\begin{eqnarray}
  \gamma_l = \left\{
    \begin{array}{@{}ll@{}}
    \dsp \left( \frac{\lambda}{4\pi d_l}\right)^2 G_t G_r 10^{-0.1\delta d_l} , & \! \!
    l = 1 \, , \\
     \dsp \beta r_l \left( \frac{\lambda}{4\pi d_l}\right)^2 G_t G_r 10^{-0.1\delta d_l} , & \! \! l=2,3,\ldots,L \, ,
  \end{array} \right.
   \label{sys_los_nlos}
\end{eqnarray}
where $d_l$ is the shortest path length of the $l$-th multipath between Alice and Bob, $\delta$ (in dB/Km) is the atmospheric absorption loss, $\beta$ is the Rayleigh roughness factor, $r_l$ is the Fresnel reflection coefficient of the $l$-th path, and $G_t,G_r$ are the antenna gains of Alice and Bob's ULA respectively given by \cite{sun2018propagation}
\vspace{-0.3cm}
\begin{align}
  G_t = N_t G_a \,, \; G_r = N_r G_a \;,
  \label{sys_G}
\end{align}
where $G_a$ is the gain of each antenna element. Note that here we consider the effect of both the free-space path loss and atmospheric attenuation loss, contrasting with prior works on THz CVQKD \cite{ottaviani2020terahertz,liu2018practical} that considered only the effect of atmospheric attenuation loss in the path loss model.

Let $\undb{H}=\undb{U} \bm{\Sigma} \undb{V}^{\dagger}$ be the SVD of the channel matrix with $\undb{U} \in \mathbb{C}^{N_r \times N_r}$ and $\undb{V} \in \mathbb{C}^{N_t \times N_t}$ being unitary matrices, and
\begin{equation}
    \bm{\Sigma} = \begin{bmatrix}
{\rm diag}\left\{\sqrt{\eta_1},\ldots,\sqrt{\eta_r} \right\} & \bm{0}_{r\times (N_t-r)} \\
\bm{0}_{(N_r-r)\times r } & \bm{0}_{(N_r-r)\times (N_t-r) }
\label{sys5}
\end{bmatrix} \, ,
\end{equation}
where $r$ is the rank of the channel matrix and $\sqrt{\eta_1},\ldots, \sqrt{\eta_r}$ are the $r$ non-zero singular values of $\undb{H}$. We consider an optimal eavesdropping attack where Eve has full control over the channel and introduces an entangled Gaussian attack. The received signal mode at Bob is then given by
\cite{jantti2017multiantenna}
\begin{equation}
    \hat{\bm{a}}_B = \undb{U}\bm{\Sigma}\undb{V}^{\dagger} \hat{\bm{a}}_A + \undb{U} \undb{S} \hat{\bm{a}}_E \, ,
    \label{sys6}
\end{equation}
where $\hat{\bm{a}}_B = [\hat{a}_{B,1},\ldots,\hat{a}_{B,N_r}]^T$ , $\hat{\bm{a}}_A = [\hat{a}_{A,1},\ldots,\hat{a}_{A,N_t} ]^T,$ denote the vectors of received mode at Bob and the transmitted mode from Alice, respectively, $\hat{\bm{a}}_E = [\hat{a}_{E,1},\ldots,\hat{a}_{E,N_t} ]^T$ denotes the vector of injected Gaussian mode by Eve, and $\undb{S}$ is a diagonal matrix given by
\begin{equation}
    \undb{S} = {\rm diag}\left\{\sqrt{1-\eta_1},\ldots,\sqrt{1-\eta_r}, \bm{1}_{(M-r)\times 1 } \right\}  \;,
\label{sys7}
\end{equation}
where $M={\rm min}(N_r,N_t)$.

\begin{figure}[h] 
\centering
\includegraphics[width=0.4\textwidth]{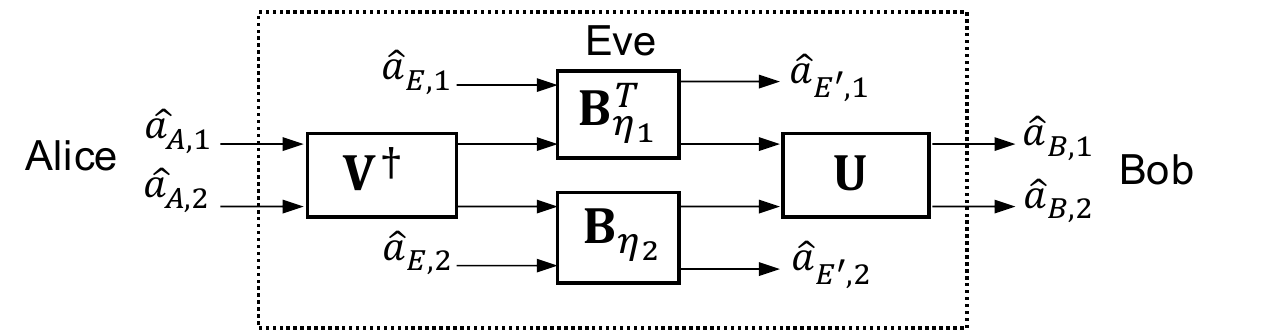}%
\caption{An illustration of beam-splitter model for a $2\times2$ MIMO channel. }
\label{fig1}
\end{figure}

To illustrate the effect of the MIMO wireless channel under consideration, we take the example of $2\times 2 $ MIMO channel which can be modelled by four beam-splitters as depicted in Fig.\ \ref{fig1}. The two signal modes transmitted by Alice are first mixed by the beam-splitter $\undb{V}^{\dagger}$, producing two output modes. These output modes are then mixed with Eve's Gaussian modes $\hat{a}_{E,1}, \hat{a}_{E,2}$ by the two beam-splitters $\undb{B}_{\eta_1}, \undb{B}_{\eta_2}$. The input-output relation of a two-port beam-splitter is given by \cite{leonhardt2003quantum}
\begin{equation}
    \begin{bmatrix}
    \hat{a}_{{\rm out},1} \\
    \hat{a}_{{\rm out},2}
    \end{bmatrix}  = \underbrace{\begin{bmatrix} \sqrt{\eta} & \sqrt{1-\eta} \\
    -\sqrt{1-\eta} & \sqrt{\eta}
    \end{bmatrix}}_{\undb{B}_{\eta}} \begin{bmatrix}
    \hat{a}_{{\rm in},1} \\
    \hat{a}_{{\rm in},2}
    \end{bmatrix} \;.
\end{equation}
Eve uses collective entanglement attack, which is the most powerful attack against Gaussian CVQKD protocols. Eve generates two pairs of two-mode squeezed vacuum states (also known as entangled Einstein-Podolsky-Rosen pair), $\{ \hat{e}_1,\hat{E}_1 \}$ and $\{ \hat{e}_2,\hat{E}_2 \}$. The first modes $\hat{e}_1, \hat{e}_2 $ are stored in quantum memory, while the other modes $\hat{E}_1, \hat{E}_2$ are mixed with the incoming mode by using $\undb{B}_{\eta_1}, \undb{B}_{\eta_2}$ respectively, as shown in Fig.\ \ref{fig1}. One output from each of the beam-splitters  ($\hat{E}'_1, \hat{E}'_2 $) are stored by Eve in her quantum memory, while the other two modes are mixed by the beam-splitter $\undb{U}$ and passed on to Bob. The ancilla modes $\{\hat{e}_1, \hat{E}'_1 \}, \{ \hat{e}_2, \hat{E}'_2 \}$ stored in Eve's quantum memory are measured by Eve after Alice and Bob finish classical communication to extract the maximum amount of information. The authors of \cite{clements2016optimal} have shown that any $M\times M$ unitary matrix can be decomposed into a mesh of interconnected two-port beam-splitters, thus, the beam-splitter model in Fig.\ \ref{fig1} can be easily generalized for a $N_r\times N_t$ MIMO model.

We assume that Alice and Bob have perfect knowledge of the channel. Alice performs transmit beamforming by using $\undb{V}$ at her end, while Bob performs receive combining by using $\undb{U}^{\dagger}$ at his end.
The effective input-output relation is thus expressed as
\begin{equation}
    \hat{\bm{a}}_B = \undb{U}^{\dagger} \undb{H} \undb{V} \hat{\bm{a}}_A + \undb{U}^{\dagger} \undb{U} \undb{S} \hat{\bm{a}}_E \, .
    \label{sys8}
\end{equation}
Using the SVD of $\undb{H}$, the effective channel between Alice and Bob is decomposed into $r$ parallel SISO channels with input-output relation given by
\begin{equation}
    \hat{a}_{B,i} = \sqrt{T_i} \hat{a}_{A,i} + \sqrt{1-T_i} \hat{a}_{E,i} \;, \quad i=1,2,\ldots,r \, ,
     \label{sys9}
\end{equation}
where $T_i$ is the $i$-th non-zero eigenvalue of $\undb{H}^{\dagger}\undb{H}$. For each of the $r$ received modes, Bob performs homodyne measurement on one of the randomly chosen quadratures. After the measurement, the input-output relation of the $i$-th parallel channel between Alice and Bob is given by a generic quantum channel with transmittance $T_i$,
\begin{equation}
 \hat{X}_{B,i}=  \sqrt{T_i} \hat{X}_{A,i} + \sqrt{1-T_i} \hat{X}_{E,i}  \;, \quad i=1,2,\ldots,r \, ,
  \label{sys10}
\end{equation}
and the input-output relation of Eve's ancilla mode is
\begin{equation}
    \hat{X}_{E',i} = -\sqrt{1-T_i} \hat{X}_{A,i}  + \sqrt{T_i} \hat{X}_{E,i} \;, \, i=1,2,\ldots,r \, ,
     \label{sys11}
\end{equation}
where $\hat{X}_{B,i} $ is the received quadrature at Bob, $\hat{X}_{A,i}$ is the transmitted quadrature by Alice, $\hat{X}_{E,i} $ is the excess noise quadrature introduced by Eve, and $\hat{X}_{E',i} $ is the ancilla quadrature stored in Eve's quantum memory for the $i$-th parallel channel. Note that the variable $\hat{X}$ corresponds to one of the two quadratures $\{ \hat{Q}, \hat{P}\}$, such that $\hat{X}_{A,i}= \{ \hat{Q}_{A,i}, \hat{P}_{A,i}\}$ and similar notation holds for $\hat{X}_{B,i}$, $\hat{X}_{E,i}$, and $\hat{X}_{E',i}$. The variance of Alice's transmitted mode is $V(\hat{X}_{A,i}) = V_a = V_s+V_0$ where $V_s$ is the variance of the initial signal encoding and $V_0$ is the  variance of the vacuum state, and $V(\hat{X}_{E,i} )=W$ is the variance of the excess noise injected by Eve. Finally, the variance of the $i$-th received mode at Bob is given by $V(\hat{X}_{B,i}) = T_i V_a+ (1-T_i) W$.


\vspace{-0.2cm}
\section{Secret Key Rate}
In the CVQKD protocol, Alice and Bob first generate a string of correlated random vectors $\{\hat{\bm{X}}_{A,n}, \hat{\bm{X}}_{B,n}\}_{n=1}^{N}$ by repeating $N$ times the quantum communication protocol described in the previous section.
Following this, Alice and Bob carry out a reconciliation protocol to correct any errors and finally perform privacy amplification to reduce Eve's knowledge of the key \cite{weedbrook2010quantum}. In this work we consider only reverse reconciliation (RR) scheme (where Bob declares on a public channel which of the two quadratures were measured by him) since RR has a higher secret key rate as compared to the direct reconciliation (DR) scheme \cite{weedbrook2010quantum,ottaviani2020terahertz}. Moreover, DR requires $T_i >0.5$ for positive secret key rates \cite{weedbrook2010quantum}, which is difficult to achieve in practice due to high path-loss and atmospheric absorption loss in the THz frequency range \cite{ottaviani2020terahertz}.

Here we generalize the secret key rate analysis of the SISO channel carried out in \cite{weedbrook2010quantum} for the MIMO channel. The asymptotic secret key rate in RR of the $i$-th parallel channel is given by \cite[Eq.~21]{weedbrook2010quantum}
\begin{equation}
    R^{\blacktriangleleft}_i = I\left( X_{A,i}:X_{B,i}\right)-I\left(X_{B,i} :E_i\right) \, ,
    i=1,\ldots,r \, ,
    \label{rate1}
\end{equation}
where $I\left( X_{A,i}:X_{B,i}\right) $ is the classical mutual information between Alice and Bob, and $I\left( X_{B,i}:E_i\right) $ is Eve's accessible information with respect to Bob's measured variables $X_{B,i}$ for the $i$-th parallel channel. The mutual information between Alice and Bob for the $i$-th pair of variables is given by
\begin{equation}
   I\left( X_{A,i}:X_{B,i}\right) = \frac{1}{2} \log_2\left[1+ \frac{T_iV_s}{\Lambda_i
   \left(V_0,W \right)}  \right] \;,
   \label{rate2}
\end{equation}
where $\Lambda_i(x,y)\dn T_ix+(1-T_i)y$, and  $V_s,V_0,W$ are the variances of the different modes defined below (\ref{sys11}).
Further, the maximum information that Eve can extract from the $i$-th parallel channel is given by the Holevo information \cite{weedbrook2010quantum} as
\bsub
\begin{eqnarray}
&& \! \! \! \! \! \! \! \! \! \! \! \! \! \!
\! \! \! \! \! \! \! \! \! \! \! \! \! \!
I\left( X_{B,i}:E_i\right) = S \left( E_i \right) - S \left( E_i \big| X_{B,i} \right) \nonumber \\
&& \quad = h \left( \nu_{1}^i \right) + h \left( \nu_{2}^i \right)
- \left( h \left( \nu_{3}^i \right) + h \left( \nu_{4}^i \right) \right) \, ,
    \label{rate3}
\end{eqnarray}
where $S(E_i)$ and $S(E_i|X_{B,i})$ denote the von Neumann entropy of Eve's state and the conditional von Neumann entropy of Eve's state given $X_{B,i}$, respectively, which are further dependent on the symplectic eigenvalues
$\nu_{k}^i \, \left( k \in \left\{ 1,2,3,4 \right\} \right)$ of the covariance matrix of Eve's state for the $i$-th parallel channel. Moreover, the function $h(x)$ in (\ref{rate3}) is defined as
\begin{equation}
    h(x) = \frac{(x+1)}{2}\log_2\frac{(x+1)}{2}- \frac{(x-1)}{2}\log_2\frac{(x-1)}{2} \;.
    \label{rate5}
\end{equation}
\esub
Here we present approximate simplified expressions for the symplectic eigenvalues in the large modulation limit; i.e., for $V_s\gg V_0,W$. The exact expression for the symplectic eigenvalues $\nu_{1}^i,\nu_{2}^i$ are given by \cite[Eq.~39]{weedbrook2010quantum} and that of $\nu_3^i,\nu_4^i$ is given by \cite[Eq.~16, 47]{weedbrook2010quantum}. These can be approximated for large $V_s$ as
\begin{align}
  \nu_{1}^i \approx \Lambda_i(W,V_a) \;,\quad
  \nu_{2}^i \approx W \;,
 \label{rate6}
\end{align}
\begin{equation}
    \nu_{3,4}^i \approx \sqrt{\frac{1}{2}\left(\Delta_i \pm \sqrt{\Delta_i^2-4\Upsilon_i } \right)} \;,
    \label{rate7}
\end{equation}
where
\bsub
\begin{equation}
    \Delta_i=\frac{V_aW\Lambda_i\left(W,V_a \right)+W\Lambda_i\left(WV_a,1 \right)}{\Lambda_i\left(V_a,W \right)} \;,
    \label{rate8}
\end{equation}
\begin{equation}
    \Upsilon_i=\frac{V_aW^2\Lambda_i\left(WV_a,1 \right)\Lambda_i\left(W,V_a \right)}{\Lambda_i^2\left(V_a,W \right)}
     \label{rate9} \;,
\end{equation}
\esub
and $V_a=V_s+V_0$.

The overall secret key rate of the MIMO CVQKD system is given by
\begin{equation}
  R^{\blacktriangleleft}_{{\rm MIMO}} =  \sum_{i=1}^{r} R^{\blacktriangleleft}_i \;.
  \label{rate_mimo}
\end{equation}
Using (\ref{rate2})-(\ref{rate9}), this may be approximated as
\beqarr
R^{\blacktriangleleft}_{{\rm MIMO}} \! \! \! \!
& \approx & \! \! \! \!
\sum_{i=1}^{r} \Bigg( \frac{1}{2} \log_2\left[1+ \frac{T_iV_s}{\Lambda_i \left( V_0,W \right)}  \right] -h(\Lambda_i(W,V_a)) \nonumber \\
&& \quad
- h(W) + h \left( \sqrt{ \frac{V_a W \Lambda_i\left( W,V_a\right)}{\Lambda_i\left( V_a,W\right)} } \right) \nn \\
&& \quad
+ h \left( \sqrt{\frac{W\Lambda_i\left( WV_a,1\right)}{\Lambda_i\left( V_a,W\right)}} \right) \Bigg) \, ,
\label{rate10}
\eeqarr
where $h(x)$ is given by (\ref{rate5}). Thus, the MIMO scheme provides a multiplexing gain of $r$, since $R^{\blacktriangleleft}_{{\rm MIMO}} $ is a sum of the secret key rates of $r$ parallel SISO channels with transmittances $T_i$.
{\color{black} In order to gain more intuition into the effect of the different system parameters on the secret key rate, we perform a Taylor series expansion of $R^{\blacktriangleleft}_{{\rm MIMO}}$ as $T_i \rightarrow 0$. (Recalling the channel model (\ref{sys1}), this is valid for $\gamma_l \to 0$, and is representative of high path loss or low signal-to-noise ratio scenarios, relevant for THz frequency applications.) This leads to  
\begin{align}
 R^{\blacktriangleleft}_{{\rm MIMO}} &\approx  \zeta \, {\rm tr}\left( \mathbf{H}^{\dagger} \mathbf{H} \right)  -rh(W)   \;,
 \label{rate11}
\end{align}
where
\begin{equation}
    \zeta= 0.72 \left[ \frac{V_s}{W} - \ln\left( \frac{V_a+1}{V_a-1}\right) \left( \frac{V_a^2-W^2}{2W}-V_a \right) \right] \;,
     \label{rate12}
\end{equation}
and where we have used $\sum_{i=1}^{r}T_i = {\rm tr} \left( \mathbf{H}^{\dagger} \mathbf{H} \right)$. One can verify from (\ref{sys1})-(\ref{sys_G}) that ${\rm tr} \left( \mathbf{H}^{\dagger} \mathbf{H} \right) \propto N_rN_t$. Thus, similar to classical MIMO communications, MIMO QKD  provides a beamforming gain of $N_rN_t$.  As seen from (\ref{rate11}), this translates to increased secret key rate by a factor of $N_rN_t$. The simulation results in the next section reveal that the approximation in (\ref{rate11}) is accurate for practical transmission distances and frequencies. We note that only the first term in (\ref{rate11}) is frequency $(f_c)$ dependent. Specifically, $\zeta$ in (\ref{rate11}) depends on $f_c$ and environment temperature $(T_e)$ since $V_a=V_s+V_0$, and $V_0=2\bar{n}+1$ with $\bar{n} = \left[\exp(hf_c/k_BT_e)-1\right]^{-1}$, where $h$ is the Planck's constant, and $k_B$ is the Boltzmann's constant. For a given $T_e$, as $f_c$ increases, $V_0$ decreases which in turn increases $\zeta$. On the other hand, the equivalent channel transmittance  ${\rm tr} \left( \mathbf{H}^{\dagger} \mathbf{H} \right)$ decreases due to a higher free-space path loss at higher frequencies. Therefore, these two competing factors have to be balanced in order to obtain a positive secret key rate. Since ${\rm tr} \left( \mathbf{H}^{\dagger} \mathbf{H} \right) >0$ for the entire frequency range, a necessary condition for positive secret key rate is $\zeta>\alpha$, where 
\begin{equation}
    \alpha= \frac{rh(W)}{{\rm tr} \left( \mathbf{H}^{\dagger} \mathbf{H} \right)} \;.
\end{equation}
Thus, the condition $\zeta>\alpha$ can be used to determine the frequency range for which secure transmission may be achieved.
\vspace{-0.3cm}
}

\begin{figure}[htp] 
\centering
\includegraphics[width=0.4\textwidth]{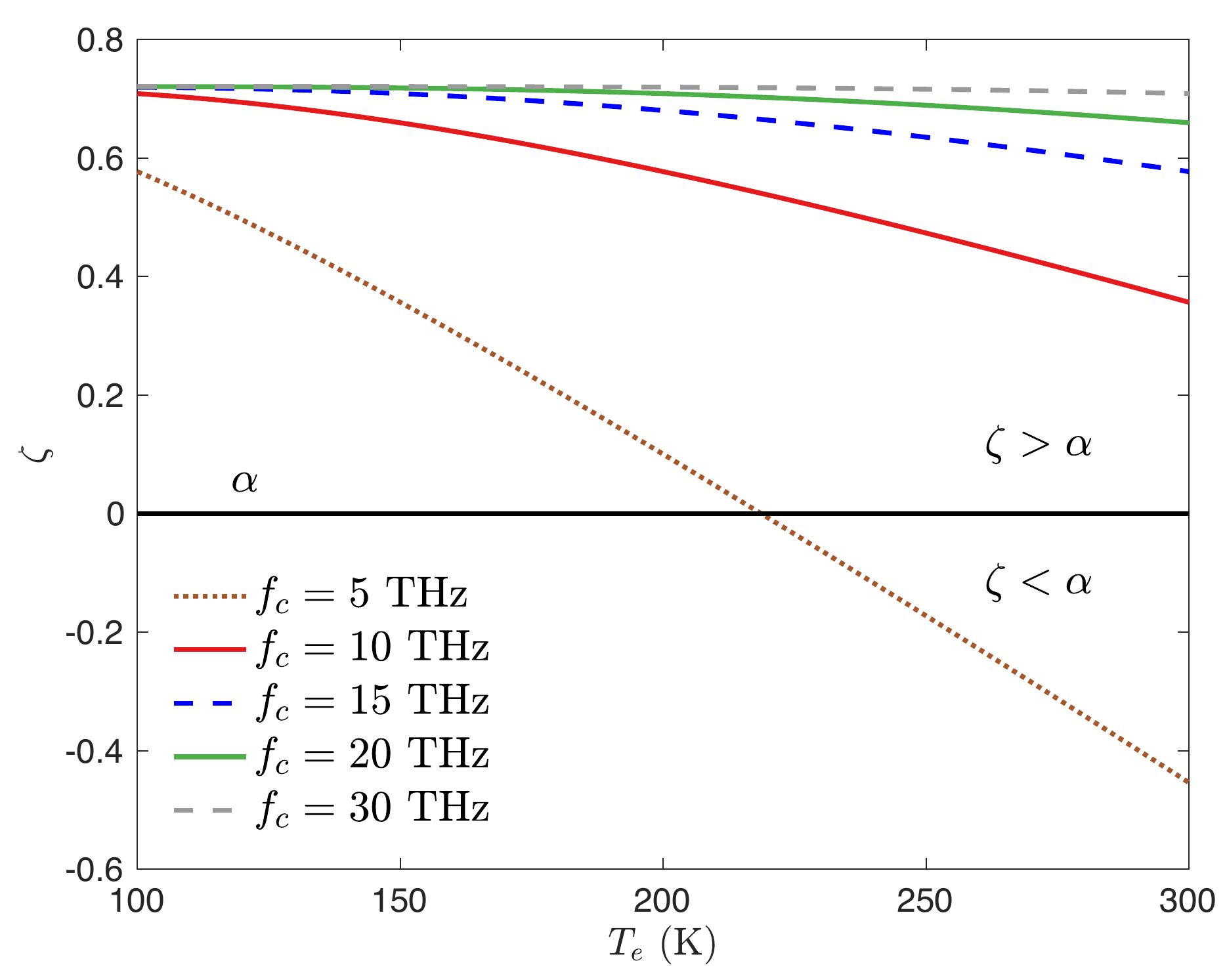}%
\caption{{\color{black} The plots show $\zeta$ from (\ref{rate12}) as a function of temperature for different frequencies with $V_s=10^3$ and $W=1$ \cite{weedbrook2010quantum}.}
}
\label{fig_coeff}
\vspace{-0.5cm}
\end{figure}

\begin{figure*}[htp] 
\centering
\subfigure[$f_c = 10$ THz ]{%
\includegraphics[width=0.33\textwidth]{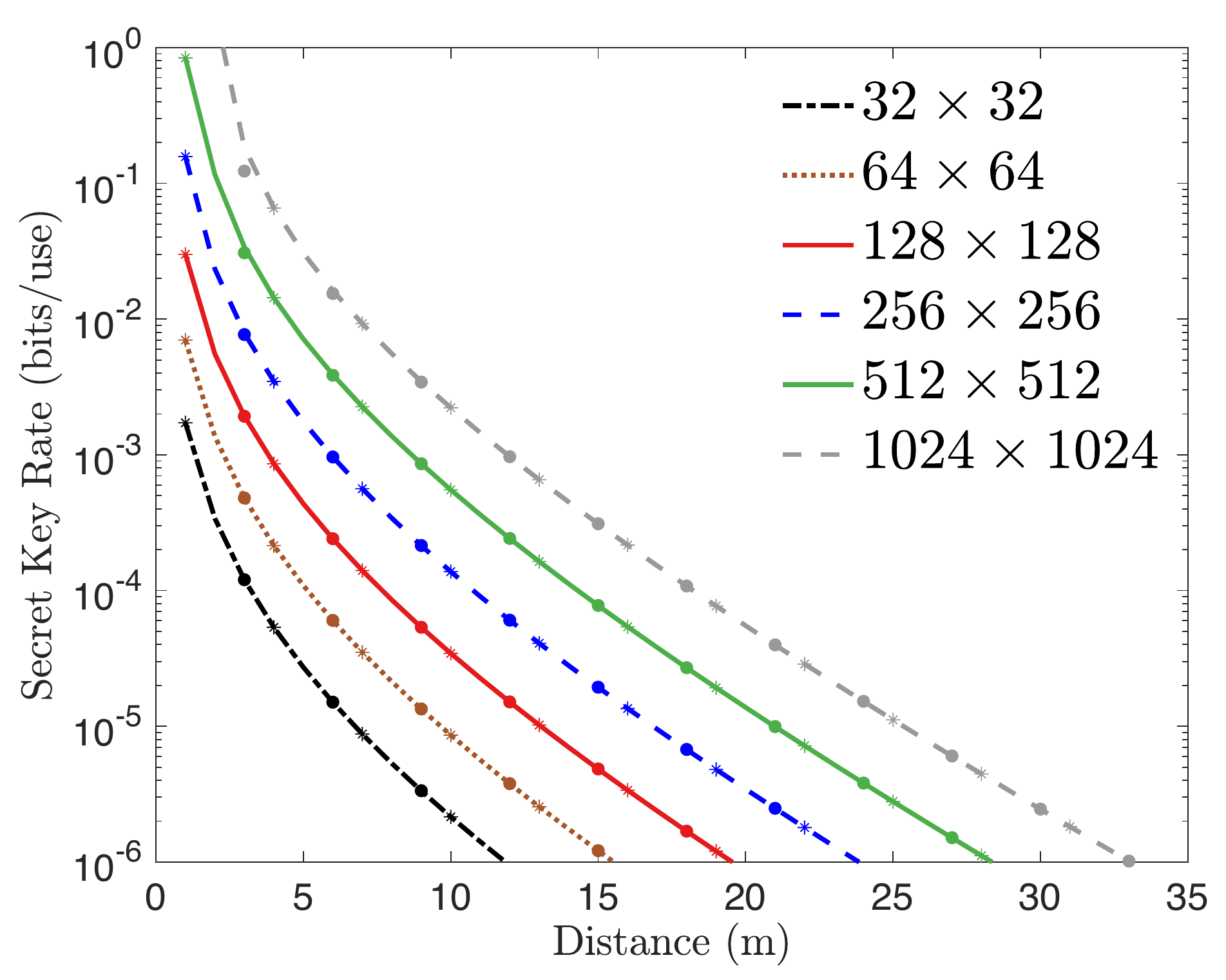}%
\label{fig_sim1:a}%
}\hfil
\subfigure[$f_c = 15$ THz  ]{%
\includegraphics[width=0.33\textwidth]{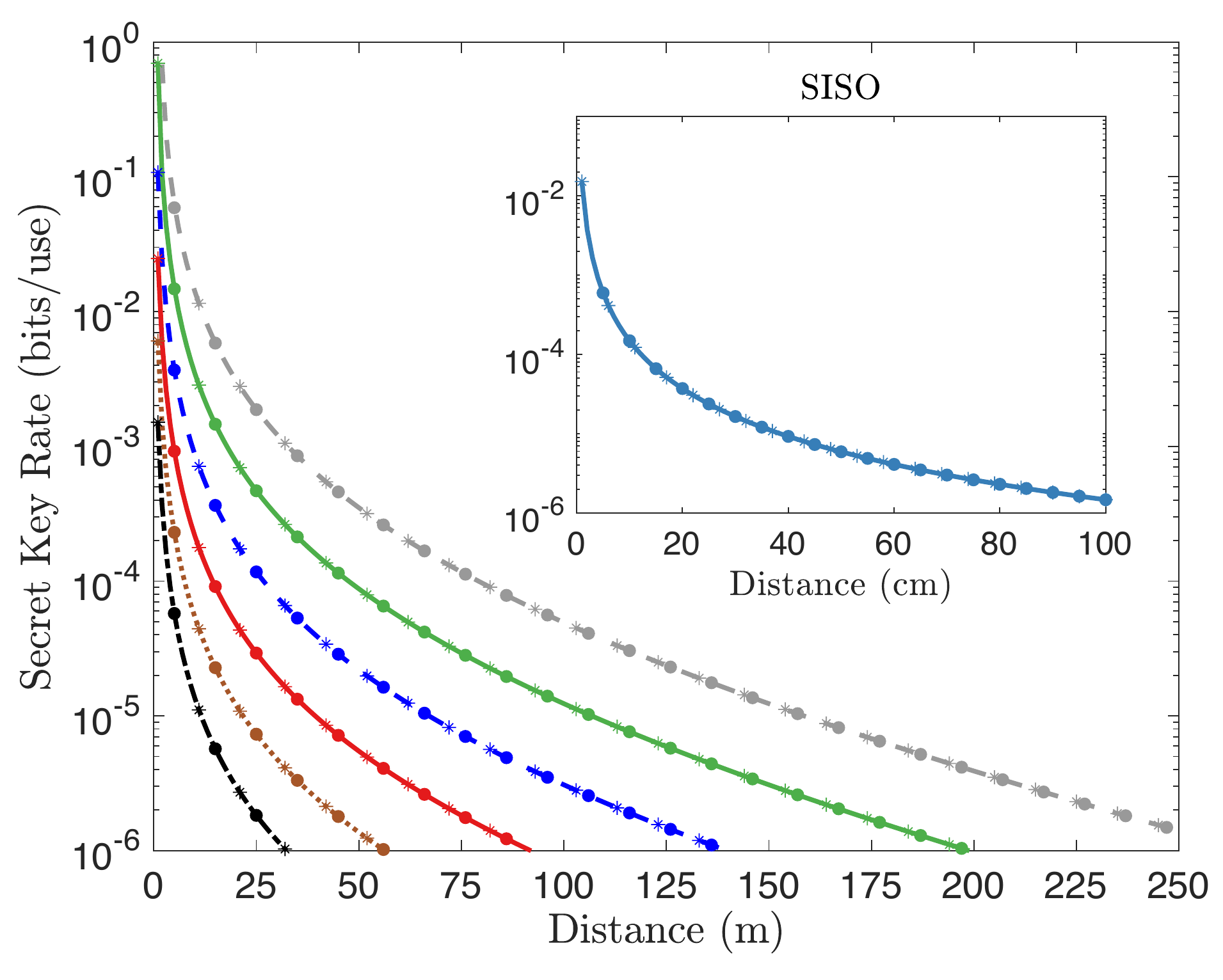}%
\label{fig_sim1:b}%
}
\subfigure[$f_c = 30$ THz  ]{%
\includegraphics[width=0.33\textwidth]{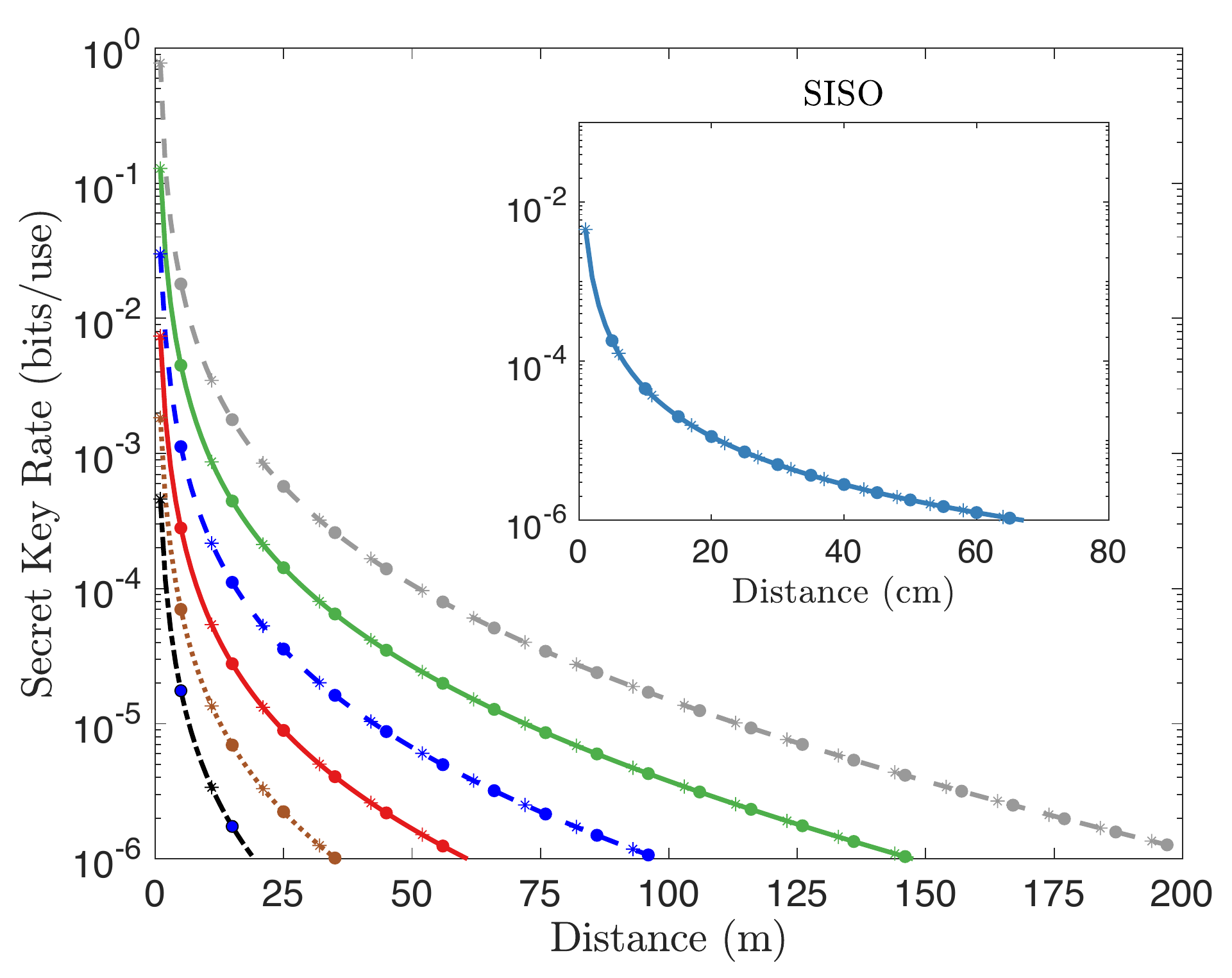}%
\label{fig_sim1:c}%
}
\caption{{\color{black}The plots show $R^{\blacktriangleleft}_{{\rm MIMO}}$ as a function of transmission distance for MIMO and SISO QKD schemes.} The asterisk represents the rate based on (\ref{rate_mimo}) with the exact expressions for the symplectic eigenvalues from \cite{weedbrook2010quantum}, the solid/dashed/dotted lines represent the large modulation approximation in (\ref{rate10}), {\color{black} and the circle represents the Taylor series expansion of (\ref{rate11})}. The simulation parameters are: $\delta=1000$ dB/Km at $10$ THz, $\delta=100$ dB/Km for $10<f_c\leq14$ THz, and $\delta=50$ dB/Km for $14<f_c\leq30$ THz \cite{sun2016predicting,ottaviani2020terahertz}, $V_s=10^3$, $W=1$, $T_e=296$ K \cite{weedbrook2010quantum}, and the antenna gain is $G_a=30$ dBi \cite{hwu2013terahertz}.
}
\label{fig_sim1}
\vspace{-0.6cm}
\end{figure*}

\begin{figure}[htp] 
\centering
\includegraphics[width=0.4\textwidth]{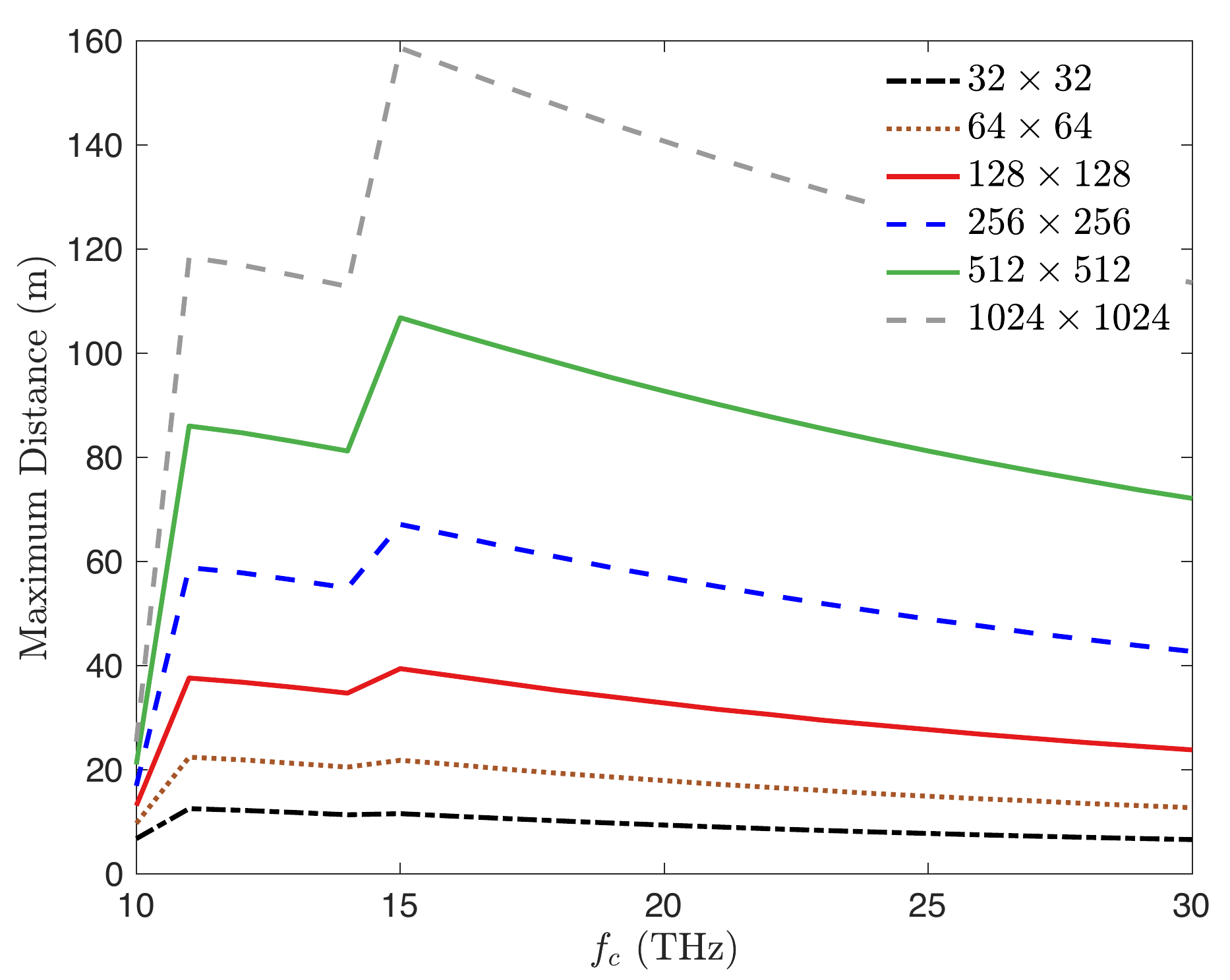}%
\caption{The plots show the maximum transmission distance as a function of frequency for a target secret key rate of $R^{\blacktriangleleft}_{{\rm MIMO}}=10^{-5}$. The other simulation parameters are the same as that of Fig. \ref{fig_sim1}.
}
\label{fig_sim2}
\vspace{-0.5cm}
\end{figure}

\vspace{-0.3cm}
\section{Simulation Results}
For our simulations, we consider a scenario with a dominated LoS path such that $L=1$. 
{\color{black} First, in Fig. \ref{fig_coeff}, we plot $\zeta$ from (\ref{rate12}) as a function of temperature (in Kelvin) for different frequencies.
It can be observed that the condition $\zeta>\alpha$ is met at room temperature ($T_e \sim 300$K) for $10\; {\rm THz} \leq f_c\leq 30\; {\rm THz}$.} Above this frequency range the atmospheric absorption and free-space path loss are particularly high, and below this frequency range the preparation thermal noise $(V_0)$ is particularly high, making it infeasible to obtain positive secret key rates at room temperature \cite{ottaviani2020terahertz}.


Fig.\ \ref{fig_sim1} shows the secret key rates $R^{\blacktriangleleft}_{{\rm MIMO}}$ (in bits/channel use) as a function of transmission distance for different MIMO configurations $(N_t \times N_r)$ at three different frequencies. For comparison, we also show the secret key rate for the SISO case at $f_c=15,30$ THz. {\color{black} The plots show that the large modulation approximation in (\ref{rate10}) and the Taylor series expansion in (\ref{rate11}) are both accurate. }
It can be observed that the secret key rates and the maximum transmission distance are significantly improved by using MIMO technology. (Note that the transmission distance is in meters for MIMO, whereas it is in centimeters for the SISO scheme.) {\color{black} Although here we show the results only for $N_r=N_t$, the same trends are also observed when $N_r \neq N_t$.}

Fig.\ \ref{fig_sim2} plots the maximum transmission distance as a
function of frequency for a target secret key rate of
$R^{\blacktriangleleft}_{{\rm MIMO}}=10^{-5}$ (bits/channel use).
We observe that the maximum transmission distance can be of the
order of $10$ m for a $32 \times 32$ MIMO, which would appear
sufficient for indoor THz QKD applications. Moreover, by using a
$1024\times 1024$ antenna array the maximum distance is around
$160$ m at $15$ THz, which is well suited for outdoor
applications. 
{\color{black} Further, it can be observed that the plots in Fig.\ \ref{fig_sim2} are not smooth and have sharp peaks. This observation can be understood from the fact that the free-space path loss monotonically increases with increasing $f_c$, while the atmospheric absorption coefficient $\delta$ varies considerably and is discontinuous in the $10-30$ THz frequency range (see the caption of Fig. \ref{fig_sim1}). 
}

\vspace{-0.2cm}
\section{Conclusions}
We have proposed a MIMO transmission scheme {\color{black}
employing SVD based transmit-receive beamforming} for THz CVQKD.
{\color{black} The multiplexing and beamforming gain obtained by using a MIMO scheme increase the secret key rate and the maximum transmission distance as compared to the SISO scheme.} Our
simulation results suggest that multiple antennas are necessary to overcome the high free-space path loss in the THz frequency range,
and $10-30$ THz is a suitable frequency window that can be used for both indoor and outdoor THz QKD applications operating at room
temperature. {\color{black} The higher secret key rate obtained
from the MIMO QKD scheme can be used to provide one-time-pad based physical layer security to high rate B5G applications. Moreover, the enhanced
maximum transmission distance can be used to
provide secure connectivity to more distant users. 
}  
In this work, we have
assumed perfect knowledge of the MIMO channel. Practical
MIMO channel estimation schemes in the quantum domain, along with
the incorporation of finite size effects and channel estimation
errors in the secret key rate analysis, need to be investigated in
future works.

\bibliographystyle{IEEEtran}
\bibliography{IEEEabrv,MIMO_QKD}

\end{document}